\documentclass[a4paper,nofootinbib]{revtex4}
\usepackage[utf8]{inputenc}
\usepackage{lipsum}
\usepackage{xcolor}
\usepackage{amssymb,amsmath,amsthm,amsfonts,mathrsfs}
\usepackage{graphicx}
\usepackage{tabularx}
\usepackage{booktabs}
\usepackage[hidelinks]{hyperref}
\usepackage[capitalise]{cleveref} 

\usepackage{xcolor}
\usepackage{subfigure}
\usepackage[normalem]{ulem}
\newcommand{\be}{\begin{equation}}
\newcommand{\ee}{\end{equation}}
\newcommand{\rH}{r_{\text{H}}}
\newcommand{\dd}{\text{d}}

\def\ii{\text{i}}

\usepackage{tikz}

\begin{document}

\title{\textbf{Nucleation of de Sitter from the anti  de Sitter spacetime in scalar field models}}

\author{M. Cadoni$^{c,d}$}
\email{mariano.cadoni@ca.infn.it}

\author{M. Pitzalis$^{c,d}$}
\email{mirko.pitzalis@ca.infn.it}

\author{A. P. Sanna${}^{c,d}$}
\email{asanna@dsf.unica.it}

\affiliation{$^{c}$Dipartimento di Fisica, Universit\`a di Cagliari, Cittadella Universitaria, 09042 Monserrato, Italy}
\affiliation{$^{d}$I.N.F.N, Sezione di Cagliari, Cittadella Universitaria, 09042 Monserrato, Italy}

\begin{abstract}

We show that, in the framework of Einstein-scalar gravity, the  gravitational coupling can drive the nucleation of the de Sitter (dS) spacetime from an anti de Sitter (AdS) one. This is done using a static and spherically-symmetric metastable scalar lump solution. This features an $\text{AdS}_4$ spacetime in the solution's core, allows for $\text{dS}_4$ vacua and is plagued by a tachyonic instability. Using the Euclidean action formalism in the semiclassical approximation, we compute and compare the probability amplitudes and the free energies of the $\text{AdS}_{4}$ lump and the $\text{dS}_4$ vacua. Our results show that the former is generally less favored than the latter, with the most preferred state being a $\text{dS}_4$ vacuum. Thus, we interpret the lump as a metastable state which mediates the decay of the $\text{AdS}_4$ spacetime into the true $\text{dS}_4$ vacuum. We argue that this nucleation mechanism of dS spacetime may provide insights into the short-distance behavior of gravity, in particular for the characterization of supergravity vacua, cosmological inflation and the black-hole singularity problem. 

\end{abstract}

\maketitle

\tableofcontents

\section{Introduction}
\label{sec:introduction}

Observational evidence~\cite{SupernovaSearchTeam:1998fmf,SupernovaCosmologyProject:1998vns,SupernovaSearchTeam:2004lze,Albrecht:2006um,SDSS:2009ocz,Planck:2018vyg} has robustly established that our universe is undergoing a phase of exponential expansion. This phenomenon, also believed to have occurred shortly after the big bang during the inflationary phase~\cite{Guth:1980zm,Linde:1981mu,Linde:2007fr,Planck:2018jri,Achucarro:2022qrl}, is well-described by the de Sitter (dS) solution within General Relativity (GR). 
A GR description is expected to fail for the primordial universe, where a UV completion of the theory is required. An unpleasant feature of most UV, quantum gravity (QG) completions proposed so far is the difficulty in accommodating stable (or even metastable) dS vacua, i.e., positive energy vacua without tachyonic excitations~\cite{Maldacena:2000mw,Witten:2001kn,Dvali:2014gua,Dvali:2017eba,Dvali:2020etd,Dine:2020vmr,Berglund:2022qsb}. In contrast, negative-energy vacua, compatible with an anti-de Sitter (AdS) spacetime, appear quite natural in several approaches to QG like, e.g., supergravity and superstring theories. Moreover, a negative value of the cosmological constant is what is normally expected as the zero-point energy of quantum fields. An intriguing way to address this issue is to explore gravity-matter models that permit transitions between AdS and dS vacua. A simple and widely-used framework for studying such vacuum transition processes, both in flat and curved spacetimes, involves a single, self-interacting scalar field theory.

In flat spacetime, the theoretical framework describing vacuum transition processes is, by now, well-understood~\cite{Coleman:1977py,Callan:1977pt}. Its extension to a curved background has been pioneered by Coleman and De Luccia (CDL)~\cite{Coleman:1980aw}, providing crucial insights into the role of gravity in such phenomena. 
However, when gravity is significant, especially in regions of large curvature, such as the dS spacetime, the physical interpretation of the phenomenon becomes more nuanced. 
This is mainly because the vacuum-transition process exhibits characteristics of both quantum tunneling and thermal phase transitions. Notably, the latter is associated with saddle-point solutions different from the CDL bounce, as discovered by Hawking and Moss (HM)~\cite{Hawking:1982ga}.

In this paper, we present a mechanism for the nucleation of a dS from an AdS spacetime, employing the standard Euclidean action formalism to describe vacuum transition processes.
We consider an explicit example in the framework of four-dimensional Einstein-scalar gravity in which the dS spacetime is nucleated out of the AdS spacetime.  Unlike the standard vacuum-transition framework, the AdS spacetime is endowed with a non-trivial scalar field and is not an exact solution of the theory, but only an approximate one.

Building on the aforementioned background on phase transitions in gravity, our approach is further inspired and motivated by the natural emergence of scalar fields in string theory compactifications, their usefulness in modeling inflation, and their role in constructing regular (singularity-free) compact object solutions that interpolate between AdS and dS spacetimes in different regimes~\cite{Franzin:2018pwz,Cadoni:2023wxa}. Furthermore, recent studies in two-dimensional dilatonic theories have shown the emergence of dS spacetime from AdS fluctuations~\cite{Biasi:2022ktq,Ecker:2022vkr}. The nucleation of a dS spacetime is also relevant in the context of regular black holes, which often exhibit cores with a dS geometry~\cite{Hayward:2005gi,Ansoldi:2008jw,Bonanno:2000ep,Cadoni:2022chn}. Finally, recent calculations in the Functional Renormalization Group framework show the emergence of such a dS core from an AdS one~\cite{Bonanno:2024wvb}. 

The transition described in this paper is mediated by regular solutions of Einstein-scalar gravity. Despite exhibiting a rich spectrum of configurations~\cite{Cadoni:2023wxa}, their vacuum structure is tightly constrained. Specifically, we demonstrate that such solutions cannot describe transitions between an exact Minkowski or (A)dS vacuum in the core to an arbitrary external, asymptotic solution. The only vacua allowed are isolated GR solutions with a constant scalar field profile. This finding contributes to the growing body of no-go theorems constraining the spectrum of regular solutions in Einstein-scalar gravity~\cite{Galtsov:2000wmw,Bronnikov:2001ah,Bronnikov:2001tv,Cadoni:2023wxa}. Regular solutions in this framework typically feature an approximate Minkowski or (A)dS behavior in the core, driven by a non-trivial scalar field and a scalar potential linear in the field. As a proof of concept, we here consider a regular lump solution firstly found in Ref.~\cite{Cadoni:2023wxa}, which interpolates between an approximate $\text{AdS}_4$ spacetime in the core and a $\text{dS}_2 \times \text{S}^2$ topology at infinity. Notably, the scalar potential that generates such solution also features stationary points where the geometry is $\text{dS}_4$, with one being an absolute stable minimum. We show that the lump instability, initially investigated in Ref.~\cite{Cadoni:2023wxa} and explored further here, makes this solution a metastable state which eventually decays into the true, stable $\text{dS}_4$ vacuum. 

The paper is structured as follows.

In~\cref{sect:2}, after a brief review of Einstein-scalar gravity, we prove, in all generality, that regular solutions within this theory do not admit exact Minkowski or (A)dS vacua in the core.

Motivated by this result, in~\cref{section:aslp} we investigate the general properties of regular scalar solutions where the scalar potential is linear in the core. In~\cref{sec:LumpSol}, we present the regular scalar lump solution, which belongs to this class. We investigate its properties, alongside with the structure of the dS$_4$ vacua. 

In~\cref{section:vtppg} we briefly review the current understanding of vacuum-to-vacuum transitions within the Euclidean action framework.

This is instrumental in the analysis of \cref{subsec:NucldS}, where we show, using standard Euclidean methods, how this lump solution can mediate the transition between the approximately AdS$_4$ spacetime in the core to the exact stable dS$_4$ vacua. We then briefly discuss the interpretation of the transition process.

Finally, we draw our conclusions in~\cref{sect:conclusions}.

\section{Existence of (anti) de Sitter interpolating solutions in Einstein-scalar gravity}
\label{sect:2}

We consider Einstein's gravity minimally coupled to a real scalar field $\phi$ \footnote{We adopt units in which $c = \hbar = 16\pi G = 1$.} 
\begin{equation}\label{action}
\mathcal{S} = \int \dd^4 x \, \sqrt{-g} \left(\mathscr{R}-\frac{1}{2}g^{\mu\nu}\partial_\mu \phi \partial_\nu \phi - V(\phi) \right)\, ,
\end{equation}
where $\mathscr{R}$ is the Ricci scalar, while $V(\phi)$ is the self-interaction potential. The field  equations read
\begin{equation}
\begin{split}
    G_{\mu\nu} &= \frac{1}{2} T_{\mu\nu}\, , \qquad \Box \phi = \frac{\dd V}{\dd \phi}\, ,\\
    T_{\mu\nu}  &= -g_{\mu\nu} \left(\frac{1}{2}\partial_\alpha \phi \partial^\alpha \phi + V\right)+\partial_\mu \phi \partial_\nu \phi\, .
\end{split}
\label{EinsteinscalarEOM}
\end{equation}
In the following, we will deal with spherically-symmetric, static solutions of the form
\begin{equation}
\begin{split}
    \dd s^2 &= -U(r) \dd t^2 + U(r)^{-1} \dd r^2 + R(r)^2 \dd \Omega^2\, , \\
    \phi &= \phi(r)\, .
\label{staticgeneralsol}
\end{split}
\end{equation}
The field equations \eqref{EinsteinscalarEOM} then read
\begin{subequations}
\begin{align}
    &\phi'' = -\left(2 \frac{R'}{R}+\frac{U'}{U} \right)\phi' + \frac{1}{U}\frac{\dd V}{\dd \phi}\, \label{eq1};\\
    &U'' = -2\frac{R'}{R}U'-V\, \label{eq2};\\
    &\frac{R''}{R}=-\frac{1}{4}\phi'^2\, \label{eq3};\\
    &U R'^2 -1 + U R R'' - \frac{U''}{2}R^2 = 0\, . \label{eq4}
\end{align}
\label{Einsteinsequationsscalarfield}
\end{subequations}
Constant scalar-field configurations correspond to standard GR solutions (vacua)
\begin{equation}
    R(r) = r \, , \qquad U_\text{GR} = 1-\frac{V(\phi_0)}{6} r^2 \, , \qquad \phi(r) = \phi_0\, .
\label{GRsolutions}
\end{equation}
\Cref{GRsolutions} is a solution of the equations~\eqref{Einsteinsequationsscalarfield} if and only if $\phi_{0}$ corresponds to the position of an extremum of the potential $V(\phi)$. Depending on the sign of $V(\phi_0)$, \cref{GRsolutions} describes Minkowski, dS or AdS vacua for $V(\phi_0) = 0$, $V(\phi_0) > 0$ or $V(\phi_0) < 0$, respectively. The (A)dS vacua are characterized by a length-scale $L$, which is related to the cosmological constant in the action \eqref{action} and is, therefore, determined by the value of the potential at the extrema: $|V(\phi_0)|/6 = L^{-2}$.  In the dS case, $L$ also gives the size of the cosmological horizon.

Apart from the standard GR ones, the theory allows also for vacua with $\text{dS}_2\times \text{S}^2$ topology, always  located at  extrema of the potential $V(\phi)$:
\begin{equation}
    R = \sqrt{\frac{2}{V(\phi_0)}} \, , \quad U_\text{GR} = 1-\frac{V(\phi_0)}{2} r^2 \, , \quad \phi(r) = \phi_0\, .
\label{ds2solutions}
\end{equation}
\Cref{GRsolutions,ds2solutions} are not the most general solutions of the theory \eqref{action} endowed with a trivial scalar field. In \cref{GRsolutions} we may include the Schwarzschild term $c_1/r$. Moreover, depending on the shape of the potential $V(\phi)$, we may have also solutions with a nontrivial $\phi(r)$ profile. In this paper, we are interested in solutions which are regular in the near $r=0$ region (the core) and have regular asymptotics at $r\to \infty$. As first observed in Ref.~\cite{Franzin:2018pwz} and analyzed in details in Ref.~\cite{Cadoni:2023wxa}, setting $c_1 = 0$ is necessary to construct smooth and regular solutions in the core. One might also consider subleading terms with respect to $r^2$ in the metric function, such as $U = 1-\alpha + \beta r + \lambda r^2 + \mathcal{O}(r^3)$. However, the $\alpha$-term produces a curvature (conical) singularity of order $\mathcal{O}(r^{-2})$, while the $\beta$-one yields a singularity of order $\mathcal{O}(r^{-1})$. Therefore, \cref{GRsolutions} represents the most general solution in the core without curvature singularities. 

One important point when dealing with regular solutions of \cref{Einsteinsequationsscalarfield} is the existence of configurations interpolating between different vacua, as described by 
\eqref{GRsolutions} and \eqref{ds2solutions}, in the regions $r = 0$ and $r \to \infty$. These configurations can appear both as isolated solutions, corresponding to extrema of the potential, or as approximate ones in the $r=0$ and $r\to \infty$ regions of an interpolating solution. In the latter case, \cref{GRsolutions,ds2solutions} provide the leading terms of the series expansion for this configuration. However, the existence of interpolating solutions with the regular behavior \eqref{GRsolutions} in the core is not generally granted. We will explore this issue in the next section \footnote{We will not consider a $\text{dS}_2\times \text{S}^2$ behavior near $r=0$. In our model, it appears only in the $r\to \infty$ region.}. 

\subsection{Nonexistence of interpolating solutions with pure (anti) de Sitter vacua in the core}

Several no-go theorems in Einstein-scalar gravity~\cite{Galtsov:2000wmw,Bronnikov:2001ah,Bronnikov:2001tv,Cadoni:2023wxa} show that solutions of \cref{EinsteinscalarEOM} cannot interpolate between a dS spacetime at $r \sim 0$ and either a Minkowski or AdS spacetime at $r \to \infty$. However, this theorems does not preclude the possibility of interpolation between two dS spacetimes, nor does it exclude the presence of AdS or Minkowski cores.

In the following, we extend these results by proving that solutions interpolating between the Minkowski or (A)dS exact vacua \eqref{GRsolutions} in the core and any arbitrary nontrivial solution outside cannot exist within the framework of the theory \eqref{action}. The only allowed vacua are isolated GR solutions with a trivial scalar field profile. 
To do so, we start by assuming that it exists a quadratic extremum for $\phi=\phi_0$ of the potential at $r \sim 0$ and work out the solution perturbatively using the field equations \eqref{Einsteinsequationsscalarfield}. 

The scalar potential we start from is of the form
\begin{equation}
    V(\phi) = \Lambda + V_1 \phi + V_2 \phi^2 + \mathcal{O}[(\phi-\phi_0)^3]\, ,
\label{scalarpotential}
\end{equation}
where $\Lambda$, $V_1$ and $V_2$ are constants. \Cref{scalarpotential} is quite general, as the linear term can always be introduced through a translation of $\phi$. While this translation results in a shift in the position of the extremum, it does not affect our overall conclusions. $V_{1, 2}$ determine the value of $\phi_0$
\begin{equation}
\label{constraintV1V2}
    \phi_0= -\frac{ V_1 }{2 V_2}\, .
\end{equation}
If we expand the metric and scalar-field solutions in power series around $r \sim 0$, \cref{Einsteinsequationsscalarfield} constrain the expansions to be (see also Ref.~\cite{Lavrelashvili:2021rxw})
\begin{subequations}
\begin{align}
    &\phi(r) = \phi_0 + \phi_2 r^2 + \phi_4 r^4 + \mathcal{O}(r^6)\, ;\\
    &U(r) = 1 + U_2 r^2 + U_4 r^4 + \mathcal{O}(r^6)\, ;\\
    &R(r) = r + R_5 r^5 + \mathcal{O}(r^7)\, ,
\end{align}
\label{expansion}
\end{subequations}
where $\phi_2$, $\phi_4$, $U_2$, $U_4$, $R_5$ are constants. $U_2 = 0$, $U_2 >0$ or $U_2 < 0$ imply a Minkowski, AdS or dS behavior near $r \sim 0$, respectively.

Plugging \cref{expansion} into \cref{eq3} and expanding near $r \sim 0$ yields
\begin{equation}
    \left(80 R_5 + 4 \phi_2^2 \right)r^2 +\mathcal{O}(r^4) = 0\, .
\end{equation}
For this to be satisfied at the order considered, we must require $\phi_2 = 2\sqrt{5}\sqrt{-R_5}$ and $R_5 \leq 0$.

From \cref{eq4}, one gets
\begin{equation}
    \left(-5 U_4 + 30 R_5\right)r^4 + \mathcal{O}(r^5) = 0\, ,
\end{equation}
from which one obtains $U_4 = 6 R_5$ at leading order.

From \cref{eq2}, one obtains
\begin{equation}
    \left(6 U_2 + \Lambda-V_2 \phi_0^2\right) + 120 R_5 r^2 + \mathcal{O}(r^4) = 0\, .
\end{equation}
This fixes, at the zeroth order, $U_2 = \left(-\Lambda +V_2 \phi_0^2\right)/6$, while, at the $\mathcal{O}(r^2)$ order, $R_5 = 0$. Therefore, the asymptotic solution \eqref{expansion} becomes
\begin{subequations}
\begin{align}
    &\phi = \phi_0 + \mathcal{O}(r^4)\\
    &U(r) = 1 +\frac{-\Lambda + V_2 \phi_0^2}{6} r^2 + \mathcal{O}(r^6)\, ;\\
    &R(r) = r + \mathcal{O}(r^7)\, .
\end{align}
\label{expr0Vquadratic1}
\end{subequations}
Using a similar procedure, it is straightforward to demonstrate that even higher-order terms are constrained to vanish by the field equations.
%
%

We thus conclude that solutions interpolating between Minkowski or (A)dS quadratic vacua in the core and other nontrivial solutions outside are not permitted. Only isolated GR vacua with a trivial scalar field are allowed.

\section{Approximate solutions with a linear potential}
\label{section:aslp}

The results of the previous section constrain the shape of the scalar-field potential in the solution core, as no potential extrema are allowed at $r = 0$. However, they do not exclude the possibility of an approximate Minkowski/(A)dS solution in the core.
For instance, solutions exhibiting AdS behavior in the core are 
already known~\cite{Franzin:2018pwz,Cadoni:2023wxa}. Consistently with our findings, in all known cases, the AdS geometry in the core is not sourced by a constant scalar field and does not correspond to any local extremum of $V$. Instead, it is sourced by a nontrivial $\phi(r)$ and is generated by a potential which typically behaves linearly in $\phi$. An important feature of these configurations is that they must necessarily be approximate solutions of the field equations. 

Let us now investigate approximate solutions with a linear potential in the core. By setting $V_2 = 0$ in \cref{scalarpotential}, we focus on the expansion at $r \sim 0$ given by \cref{expansion}. Substituting it into the field equations \eqref{Einsteinsequationsscalarfield} yields
\begin{subequations}
\begin{align}
    \phi(r) &= \phi_0 + 2\sqrt{-5R_5} \, r^2 + \frac{\sqrt{-5R_5}\Lambda-60 R_5 \phi_0}{6} r^4 + \mathcal{O}(r^6)\, ;\\
    U(r) &= 1-\frac{\Lambda + V_1 \phi_0}{6}r^2 + 6 R_5 r^4 +\mathcal{O}(r^6)\, ;\\
    R(r) &= r + R_5 r^5 + \mathcal{O}(r^7)\, ,
\end{align}
\label{approximatelinearsolution}
\end{subequations}
together with the requirement $R_5 \leq 0$. 
Unlike the case discussed in the previous section, these approximate solutions do not correspond to extrema of the potential. Therefore, their existence is allowed in global solutions that interpolate between Minkowski or (A)dS geometries in the core and nontrivial configurations at spatial infinity, compatibly with the no-go theorem established in~\cite{Cadoni:2023wxa}. 

We conclude this section by highlighting the relevance of linear potentials from various perspectives. Firstly, they arise in axion inflationary scenarios within string theory, specifically in axion monodromy inflation. The latter arises from the compactification of a D$5$-brane in type IIB string theory. In four dimensions and for large values of the axion, this setup produces a linear scalar potential~\cite{McAllister:2008hb,Takahashi:2010ky,Baumann:2014nda}. Additionally, linear inflation also naturally emerges as a solution in Coleman-Weinberg inflation~\cite{Coleman:1973jx}, provided the inflaton has a nonminimal coupling to gravity and the Planck scale is dynamically generated~\cite{Kannike:2015kda}. 
The most intriguing application of Einstein-scalar gravity with an approximate linear potential, which we will focus on in the remainder of this paper, is related to their use in describing the cores of compact objects. The cores of several nonsingular scalar solutions, such as the Sine-Gordon solitonic configuration analyzed in Ref.~\cite{Franzin:2018pwz}, or the regular lump solution found in Ref.~\cite{Cadoni:2023wxa}, are described by an AdS spacetime sourced by an approximate linear potential. These solutions typically exhibit dynamical instabilities, most commonly tachyonic ones, although the instability timescale could also be quite long. Consequently, these solutions can be considered metastable states which eventually tunnel to a stable vacuum of the theory. The primary application we envision is a tunnelling process between an AdS and a dS spacetime, mediated by an unstable solution of the kind described above. This nucleation of dS out of AdS is potentially very interesting for several applications, including characterization of string theory vacua, inflation and the black-hole singularity problem. 

In the following, we will consider the lump solution of Ref.~\cite{Cadoni:2023wxa} as a specific model to describe this nucleation. However, we expect this to be just a particular case of a quite general class of Einstein-scalar gravity models.    

\subsection{Scalar lump solution}
\label{sec:LumpSol}

The regular lump spacetime of Ref.~\cite{Cadoni:2023wxa} is of the form \eqref{staticgeneralsol}, with
\begin{subequations}
\begin{align}
    R(r) &= \frac{r}{\left[1+\left(\frac{r}{\ell} \right)^4 \right]^{1/4}}\label{Rlump}\, ;\\
    U(r) &= \frac{c_2 r^2 \ell ^4-r^4+\ell ^4}{\ell ^2 \sqrt{r^4+\ell ^4}} \label{Usollump};\\
    \phi(r) & = \sqrt{5} \, \tan^{-1}\left(\frac{\ell^2}{r^2} \right) \label{scalarfieldlump}\, ;\\
    V(r) & = \frac{2 \left(5 c_2 r^4 \ell ^8-3 c_2 \ell ^{12}+r^{10}+15 r^2 \ell ^8\right)}{\ell ^2 \left(r^4+\ell ^4\right)^{5/2}} \label{Vlump}\, ,
\end{align}
\label{lumpfullsolution}
\end{subequations}
where $\ell$ is an arbitrary length-scale characterizing the potential $V$, and $c_2$ an integration constant that must be positive to ensure the metric signature does not change anywhere (see Ref.~\cite{Cadoni:2023wxa} for details). 
\begin{figure}[!ht]
    \centering
\includegraphics[width=0.65\linewidth]{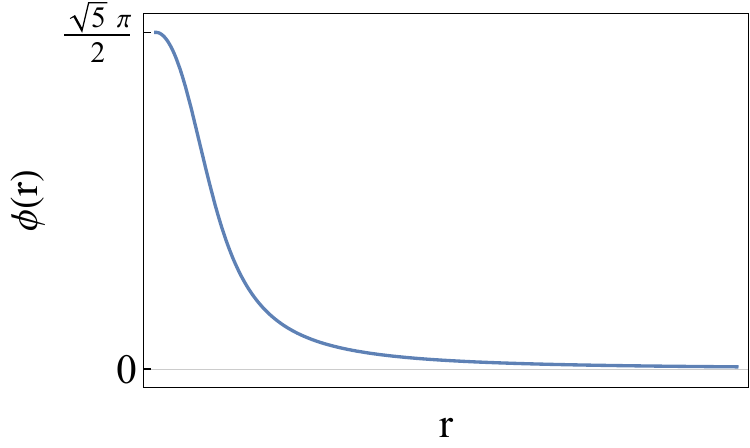}
    \caption{Qualitative behavior of the scalar field $\phi$ as a function of $r$, given by \cref{scalarfieldlump}. We have only plotted the scalar-field profile in the region of interest $r \in [0, +\infty)$. }
    \label{Fig:phiLump}
\end{figure}
Near $r\sim 0$, $R(r) \sim r$, while at infinity, $R(r) \sim \ell = \text{constant}$. $U$, instead, exhibits a dS asymptotics at infinity, with a dS length $\ell$, while near $r \sim 0$, the spacetime is AdS, with $c_2$ giving the inverse of the square of the AdS length. 

Expanding the solution \eqref{lumpfullsolution} near $r=0$ and the potential \eqref{potentialphi} near $\phi=0$, one easily finds that the lump behaves in the core exactly as the approximate AdS solution \eqref{approximatelinearsolution} with the linear potential described in the previous section. 
Therefore, the lump geometry interpolates between $\text{AdS}_4$ spacetime at $r \sim 0$ and a Nariai spacetime $\text{dS}_2 \times \text{S}^2$ at $r \to \infty$. 

The scalar field, instead, is an even function of $r$, which is limited between $\phi(0) = \sqrt{5}\pi/2$ and $\phi(\pm \infty) = 0$ (see \cref{Fig:phiLump}). \\

\begin{figure}[!t]
    \centering
\includegraphics[width=0.65\linewidth]{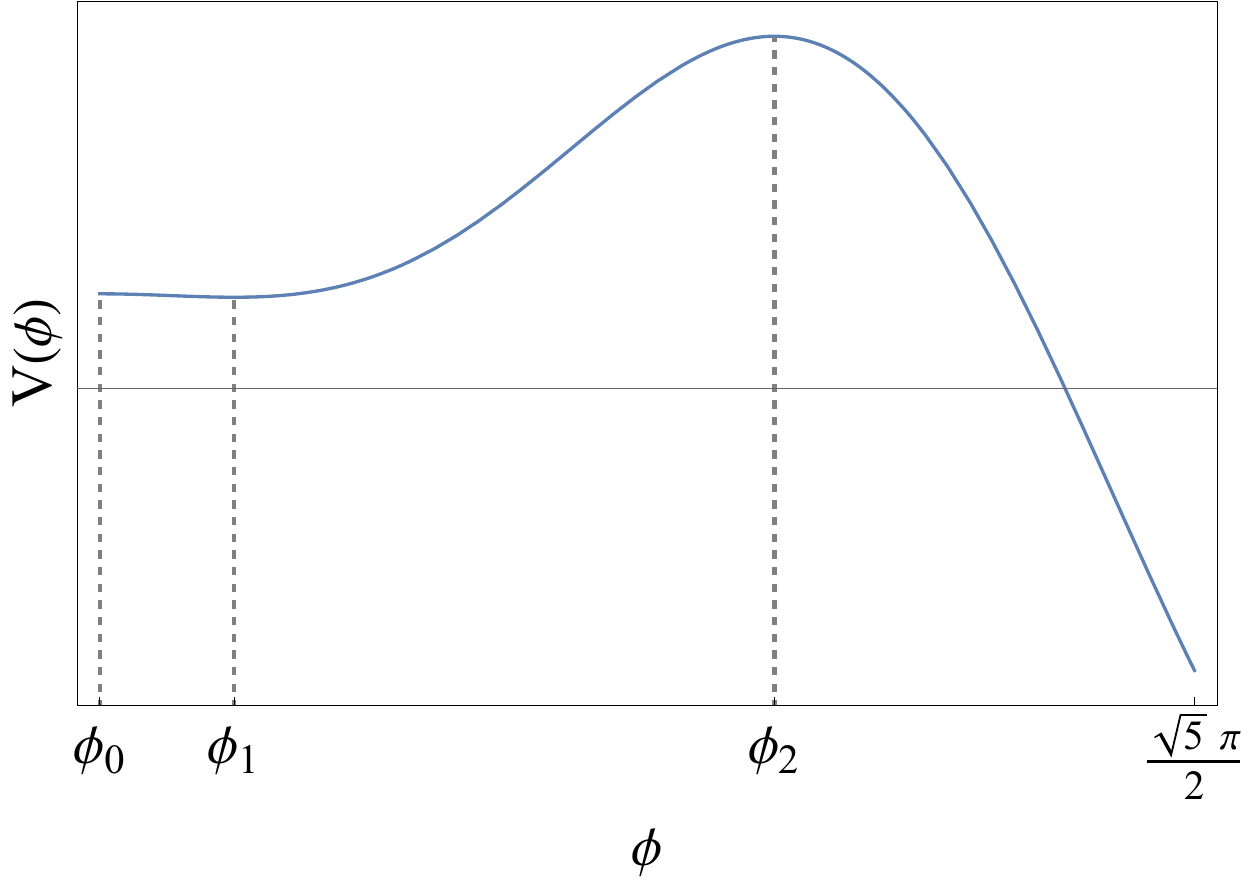}
    \caption{Qualitative behavior of the scalar lump potential as a function of $\phi$. The right side of the horizontal axis corresponds to the $r \to 0$ region, where $\phi(0) = \sqrt{5} \pi/2$, while the left side corresponds to the $r \to \infty$ region. The potential $V(\phi)$ is given in $1/\ell^{-2}$ units. The vertical dashed lines indicate the positions of the extrema, corresponding to $\text{dS}_4$ vacua. We stress that this qualitative behavior holds true for any value of $c_2$, even though we focus on the specific case $c_2 = \ell^{-2}$ in the text for simplicity.} 
    \label{Fig:VphiLump}
\end{figure}

By inverting \cref{scalarfieldlump} we can also express the potential as a function of $\phi$. To do this, we must restrict to the interval $r \in [0, \infty)$, so that $\phi(r)$ is a single-valued function. $V(\phi)$, hence, reads (see \cref{Fig:VphiLump} for a qualitative plot)
\begin{equation}
\begin{split}
    V(\phi) = &\frac{2}{\ell^2} \biggl\{c_2 \ell ^2 \left[5 \cot ^2\left(\frac{\phi }{\sqrt{5}}\right)-3\right]+\cot \left(\frac{\phi }{\sqrt{5}}\right) \left[\cot ^4\left(\frac{\phi }{\sqrt{5}}\right)+15\right]\biggr\} \sin^5 \left(\frac{\phi }{\sqrt{5}} \right)\, .
\label{potentialphi}
\end{split}
\end{equation}
Since the value of $c_2$ does not alter the relevant physical results, we will, for simplicity, set $c_2 = \ell^{-2}$ in the remainder of the paper. \\

Finally, owing to the dS asymptotic behavior, the lump solution \eqref{lumpfullsolution} features a cosmological horizon located at $U(\rH) = 0$, 
\begin{equation}\label{rHell}
\rH = \frac{\ell}{\sqrt{2}}\sqrt{1+\sqrt{5}}\, .
\end{equation}
As it will be instrumental for the computations of \cref{subsec:NucldS}, we also consider the temperature associated with this cosmological horizon $T_\text{H} = -\frac{U'(\rH)}{4\pi}$, which, using \cref{Usollump} and $c_2 = \ell^{-2}$ yields
\begin{equation}
\label{temperaturelump}
    T_\text{H} = \frac{\sqrt[4]{5}}{2\pi \ell}\, .
\end{equation}

Let us end this section by commenting two particular limits of the lump solution, i.e., $\ell \to 0$ and $\ell \to \infty$. 

By taking the limit $\ell \to 0$ of \cref{lumpfullsolution}, we obtain a $\text{dS}_2 \times \text{S}^2$ metric, with dS length $\ell$ and zero scalar field. Conversely, in the limit $\ell \to \infty$, the metric of the lump reduces to Minkowski spacetime with $\phi = \pi/2$. This is expected, as the cosmological spherical horizon gets pushed to infinity in this limit, and thus it reduces to a plane. From being compact, the manifold becomes non-compact in this limit.

We will come back to these limits in \cref{subsec:NucldS}.

\subsubsection{de Sitter vacua}
\label{sec:dSvacua}

As illustrated in \cref{Fig:VphiLump}, the potential \eqref{potentialphi}, apart from the exact lump solution \eqref{lumpfullsolution}, allows for exact GR vacua. In the interval $\phi \in [0,\sqrt{5}\, \pi/2]$, they correspond to the three extrema $\phi_0 = 0$, $\phi_1$ and $\phi_2$ (whose values depend on $c_2$). Depending on the chosen spacetime topology, they correspond either to pure $\text{dS}_4$ (see \cref{GRsolutions}) or $\text{dS}_2 \times \text{S}^2$ (see \cref{{ds2solutions}}) vacua. Notice that the topology of the vacuum $\phi_0$ is fixed to be $\text{dS}_2 \times \text{S}^2$ if it is not isolated, but generated in the $r\to \infty$ region by the interpolating lump solution \eqref{lumpfullsolution}. Conversely, there is no natural topology choice for $\phi_{1,2}$-vacua. 

The position of the extrema and the corresponding values of the potential computed at these points are as follows (with $c_2 = \ell^{-2}$)
\begin{align}
    \phi_0 &= 0 \, ,  &&V(\phi_0) = \frac{2}{\ell^2}\, ;\nonumber\\ 
    \phi_1 & \simeq 0.430664\, ,  &&V(\phi_1) \simeq \frac{1.92736}{\ell^2}\, ; \nonumber \\
    \phi_2 & \simeq 2.16417\, , && V(\phi_2) \simeq \frac{7.4697}{\ell^2}\, .
\label{dSVacua}
\end{align}
Their dS horizons are given by 
\begin{equation}
    r^{(i)}_{\text{H, dS}} = \sqrt{\frac{6}{V(\phi_i)}}\, , \qquad i = 0, 1, 2\, ,
    \label{rHpuredS}
\end{equation}
with the associated temperatures
\begin{equation}
    T^{(i)}_{\text{H, dS}} = -\frac{U'_\text{dS}\left(r^{(i)}_{\text{H, dS}}\right)}{4\pi} = \frac{1}{2\pi} \sqrt{\frac{V(\phi_i)}{6}}\, .
\end{equation}
\section{Vacuum transition processes in presence of gravity}
\label{section:vtppg}

Before going into the details of the AdS$\to$ dS transition in our model, we briefly review some general features of vacuum transitions in the Euclidean gravity framework.

In flat spacetime~\cite{Coleman:1977py,Callan:1977pt}, one usually considers a potential $V(\phi)$ featuring two, non-equivalent local minima: a false vacuum $\phi_{\text{fv}}$, with higher energy, and a true vacuum $\phi_\text{tv}$, with lower energy, i.e., $V(\phi_{\text{fv}}) > V(\phi_\text{tv})$, separated by a local maximum. The transition from the false to the true vacuum occurs via quantum tunneling through the potential barrier, with a probability amplitude characterized by $\Gamma = A \text{e}^{-B}$. In the WKB approximation, the exponent $B = \mathcal{S}_{\text{E}}(\phi)-\mathcal{S}_{\text{E}} (\phi_{\text{fv}})$ corresponds to the on-shell Euclidean action $\mathcal{S}_{\text{E}}$ of the bounce solution, a localized instanton that mediates the transition. Specifically, $\phi$ is a solution of the equation of motion with boundary conditions satisfying $\phi \to \phi_{\text{fv}}$ at spatial infinity and ensuring the instanton regularity at $r = 0$.  

The bounce solution minimizing the action and satisfying these properties exhibits an $O(4)$ symmetry~\cite{Coleman:1977th}. The pre-exponential factor $A$, instead, can be computed through path-integral methods, by taking the imaginary part of the gaussian integral over second-order variations around the classical bounce solution. Only when the second variational derivative of the Euclidean action at the bounce has one and only negative eigenvalue, however, there is a natural physical interpretation in terms of a tunneling transition between vacua~\cite{Coleman:1987rm}. 

Bounce solutions which describe more than one negative mode have an Euclidean action which is generally larger than the usual bounce, thus giving rise to subdominant contributions to the vacuum decay process (they  specify the path connecting thermally excited configurations, oscillating between the two sides of the potential barrier). 

The vacuum transition can be intuitively thought as the nucleation of true-vacuum bubbles inside the false vacuum. The inside of the bubbles, which appear as a ball of true vacuum, is separated from the outside, a sea of false vacuum, by a transition region, referred to as ``the wall". A crucial assumption in the calculation is that the thickness of the wall must be small compared to the bubble radius, or, equivalently, the energy difference between vacua must be small with respect to the height of the potential barrier~\cite{Coleman:1977py}. This is the so-called ``thin-wall" approximation. When the latter is valid, once the bubble materializes at rest with a certain radius, it grows from a (typical) subnuclear scale to a macroscopic one, accelerating to the speed of light. By energy conservation, all the energy released by the transition from false to true vacuum goes into accelerating the bubble wall (in agreement with the equations of motion), leaving behind only the true vacuum. 

This framework can be extended to encompass quantum field theory in flat spacetime at finite temperature $T$. In the high-temperature regime ($1/T$ smaller than the characteristic radius of the bounce), vacuum transition happens via bubble nucleation due to thermal fluctuations, rather than to quantum tunneling (see, e.g., Refs.~\cite{Langer:1967ax,Linde:1981zj,Quiros:1999jp,Batra:2006rz,Brown:2007sd,Weinberg:2012pjx,Lee:2014uza, Hackworth:2004xb}).

The extension to curved background has been pioneered by Coleman and De Luccia (CDL)~\cite{Coleman:1980aw}. Despite significant progress, the prescription for evaluating the pre-exponential factor $A$ in the transition probability amplitude remains a topic of debate, particularly concerning the existence and implications of single or multiple negative modes~\cite{Lavrelashvili:1985vn,Bousso:1998ed,Tanaka:1992zw,Lavrelashvili:1999sr,Khvedelidze:2000cp,Gratton:2000fj,Hackworth:2004xb,Lavrelashvili:2006cv,Lee:2011ms,Lee:2012qv,Battarra:2012vu,Lee:2014uza,Lee:2014ula,Gregory:2020cvy}. A single negative mode confirms the interpretation of the instanton as the object mediating the tunneling between vacua, while multiple ones could suggest that the instanton is, rather, a thermal intermediate configuration~\cite{Battarra:2012vu}. In contrast, the semiclassical calculation of the exponent $B$ is relatively straightforward and follows a methodology similar to the flat-spacetime case. The exponent is determined by the on-shell Euclidean action, modified to account for gravitational effects. This requires evaluating the action on a solution to Einstein's equations coupled with a self-interacting scalar field. These solutions describe the evolution of a true vacuum bubble in the presence of gravity. The geometry of the bubble is again assumed to exhibit $O(4)$ symmetry. Unlike in flat spacetime, this is not rigorously proven on a curved manifold. Nevertheless, it is generally assumed that the inclusion of gravity preserves the symmetry observed in flat spacetime, while also providing a substantial simplification in the calculations.

When gravity is included, however, the corrections induced by spacetime curvature must be carefully considered in the analysis. For small spacetime curvatures, the CDL instanton closely resembles its flat-spacetime counterpart, making tunneling possible when the potential barrier is both narrow and high \footnote{Some criticism has been raised concerning the use of the thin-shell approximation in the case of gravity, as gravitational backreaction effects may be neglected~\cite{Copsey:2011zj}. Taking into account them could imply an enhancement in the decay rate, with also possible impacts on the string theory landscape.}. However, in regions of large curvature, such as dS spacetime, vacuum transitions feature aspects of both quantum tunneling and thermal fluctuations, further complicating the interpretation of the phenomenon. In particular, an additional saddle solution arises. The latter, first identified by HM~\cite{Hawking:1982ga}, is not interpreted as mediating a vacuum transition, but rather as a thermal fluctuation of a system with a size and temperature equal to the dS ones~\cite{Rubakov:1999ir,Brown:2007sd}. They first observed that CDL bounce solutions were absent in the case of very flat potentials. This suggested the presence of an $O(5)$-symmetric static solution, localized at the top of the potential barrier—a characteristic commonly associated with bounces in thermal systems. This is the only instanton existing when the dS curvature exceeds the scalar field's effective mass near the potential maximum. In this case the solution develops an infinitely-thick wall~\cite{Jensen:1983ac,Jensen:1988zx,Rubakov:1999ir,Kanno:2011vm,Battarra:2012vu,Weinberg:2012pjx}. Consequently, the CDL instanton does not exist anymore and vacuum transition no longer involves bubble nucleation. The field, instead, undergoes a smooth transition from the top of the potential barrier to the true vacuum (for further insights, see, e.g., Refs.~\cite{Linde:1991sk,Rubakov:1999ir,Batra:2006rz,Lee:2008hz,Weinberg:2012pjx,Battarra:2013rba,Blanco-Pillado:2019xny,Espinosa:2021tgx,Camargo-Molina:2022ord,Camargo-Molina:2022paw,Miyachi:2023fss}). \\

\section{Nucleation of $\text{dS}_4$ from $\text{AdS}_4$  in the Euclidean action formalism} 
\label{subsec:NucldS}

Let us now discuss the vacuum transition process in our scalar-gravity model, with potential \eqref{potentialphi}. We begin by revisiting the tachyonic instability featured by the lump solution~\eqref{lumpfullsolution}, previously identified in Ref.~\cite{Cadoni:2023wxa}.

\subsection{Scalar tachyonic instability}
A weak condition on the presence of a tachyonic instability can be inferred by performing a linear stability analysis of scalar perturbations $\delta \phi$ around the maximum $\phi_2$, corresponding to a dS spacetime. At linear order, scalar perturbations decouple from the metric and their dynamics is described by the Klein-Gordon equation
\begin{equation}
 \Box \delta \phi = \frac{\dd^2 V}{\dd \phi^2} \biggr|_{\phi= \phi_2}  \delta \phi\, . 
 \label{deltaphiequationofmotion}
\end{equation}
The second derivative of the potential, evaluated at the maximum, can be interpreted as an effective mass squared of the perturbation, i.e., $m^2_{\delta \phi} = \partial^2_\phi V|_{\phi \, = \, \phi_2}$. Since we are considering the local behavior around the potential maximum, any small perturbation is expected to drive the system away from $\phi = \phi_2$, leading to linear instability. The latter is signaled by the negative effective squared mass of the scalar perturbation $m^2_{\delta \phi} < 0$, characteristic of a tachyonic mode. For the lump solution \eqref{lumpfullsolution}, using the potential \eqref{potentialphi} and adopting, as always, $c_2 = \ell^{-2}$, together with $\phi_2$ given by \cref{dSVacua}, we obtain $m^2_{\delta \phi} \simeq -16.2 \, \ell^{-2}$. 

The timescale of the instability can, instead, be estimated from the solution of \cref{deltaphiequationofmotion}. The calculation is most conveniently performed in the flat slicing of dS spacetime, where the metric reads as
\begin{equation}
    \dd s^2 = \frac{L^2_2}{\eta^2} \left(-\dd \eta^2 + \dd r^2 + r^2 \dd \Omega^2 \right)\, ,
\end{equation}
where $L_2$ is the dS length associated to the $\phi = \phi_2$ solution, while $\eta$ is the conformal time, related to the cosmological time $t$ by $\eta \sim e^{-t/L_2}$. The Klein-Gordon equation \eqref{deltaphiequationofmotion} in this coordinate system takes the form
\begin{equation}
    \left[\frac{\partial^2}{\partial \eta^2}-\frac{2}{\eta}\frac{\partial}{\partial \eta} -\nabla^2 +\frac{m^2_{\delta \phi} L_2^2}{\eta^2}\right]\delta \phi = 0\, ,
    \label{KGequationindS1}
\end{equation}
with $\nabla$ the spatial Laplacian operator. Given the spherical-symmetry of the system, we Fourier transform the spatial profile of $\delta \phi$, i.e., $\delta \phi(\eta, r) = \int \dd^3 k \, e^{\ii \vec{k} \cdot \vec{r}} \delta \phi_k(\eta)$, which allows us to treat the perturbation as a superposition of independent plane waves. In the long-wavelength limit, i.e., $k \to 0$, $\nabla^2 \delta \phi$ in \cref{deltaphiequationofmotion} can be neglected, and the solution reads
\begin{equation}
    \delta \phi_k(\eta) \simeq \eta^{3/2}\left[\mathcal{C}_1 \, \eta^{-\frac{L_2}{2}\sqrt{\mu^2} \, m_{\delta \phi}} + \mathcal{C}_2 \, \eta^{\frac{L_2}{2}\sqrt{\mu^2}\, m_{\delta \phi}}\right]\, ,
\label{deltaphisol}
\end{equation}
where $\mathcal{C}_{1,\, 2}$ are integration constants, and we define
\begin{equation}
    \mu^2 \equiv -4 + \frac{9}{L_2^2 \, m^2_{\delta \phi}}\, .
\end{equation}
Since $m^2_{\delta \phi} < 0$, $\mu^2<0$. Therefore, considering also $m_{\delta \phi} = \ii |m_{\delta \phi}|$ and $\mu = \ii |\mu|$, \cref{deltaphisol} becomes
\begin{equation}
    \delta \phi_k(\eta) \simeq \eta^{3/2}\left[\mathcal{C}_1 \, \eta^{ \frac{L_2}{2}|\mu| \, |m_{\delta \phi}|} + \mathcal{C}_2 \, \eta^{-\frac{L_2}{2}|\mu| \, |m_{\delta \phi}|}\right]\, .
\end{equation}
Converting to cosmological time, at late times $t \to \infty$, only the growing modes survive
\begin{equation}
    \delta \phi_k(t) \sim e^{\gamma t}\, ,
\end{equation}
with 
\begin{equation}
    \gamma \equiv -\frac{3}{2L_2} + \frac{|\mu|}{2}\, |m_{\delta \phi}|\, .
\end{equation}
The latter represents the inverse of the instability time-scale. Combining \cref{dSVacua} with the fact that $|V(\phi_2)| = 6/L_2^2$, and using $|m_{\delta \phi}|\simeq 4.02 \, \ell^{-1}$, yields $\gamma \simeq 1.99 \, \ell^{-1}$, giving a time scale $\tau_{\text{inst}}\sim 0.50 \, \ell$. 

$\tau_\text{inst}$ can, thus, be very long, comparable with the Hubble time at the formation of the lump. In the particular limit $\ell \to \infty$, $\tau_\text{inst}$ becomes infinite, meaning that the lump never decays. This is consistent with the stability of Minkowski spacetime, which is recovered in this limit from \cref{lumpfullsolution}, as already discussed at the beginning of \cref{sec:LumpSol}. 

\subsection{Vacuum transition and the nucleation of $\text{dS}_4$}

According to the results of the previous section, the lump can be regarded as a metastable state, representing a decay mode of the $\text{AdS}_4$ core-approximate solution \eqref{approximatelinearsolution} in the linear potential model. The presence of the $\text{dS}_4$/$\text{dS}_2 \times \text{S}^2$ vacua in the model suggests that the scalar lump will decay into one of the latter. At the semiclassical level, the most favored vacuum can be determined in two ways, both of which can be implemented using an Euclidean-action formalism. The simplicity of the model allows for exact analytical calculations of the Euclidean action, without the need for any approximations.

The first way is a thermodynamical one, consisting in computing and comparing the free energies of the solutions. This method is based on the semiclassical correspondence between the Euclidean action and the partition function in the canonical ensemble~\cite{Gibbons:1976ue}
\begin{equation}
    \mathcal{S}_\text{E} = -\ln \mathcal{Z} = \beta F\, .    \label{EuclideanActionPartitionFunction}
\end{equation}
$\beta$ is the inverse temperature of the ensemble, while $F$ the free energy. $\mathcal{S}_\text{E}$ has to be computed on shell, for a given classical configuration representing the Gibbons saddle of the path integral, corresponding to a classical Euclidean instanton. The most favored vacuum is the one with the least free energy.

The second method, instead, is based on computing the relative probability amplitude of the different vacuum configurations $\Gamma \sim \text{e}^{-B}$, with $B$ is determined by the on-shell Euclidean action~\cite{Coleman:1980aw,Jensen:1983ac,Hackworth:2004xb,Brown:2007sd,Weinberg:2012pjx,Gregory:2013hja,Oshita:2016oqn}, namely
\begin{equation}
    \Gamma \, \sim \, \text{e}^{-\left(\mathcal{S}_\text{E}-\mathcal{S}_\text{b}\right)}\, ,
    \label{instantonamplitude}
\end{equation}
where
\begin{equation}
\begin{split}
    \mathcal{S}_\text{E} &= -\int \dd^4 x \, \sqrt{g_\text{E}} \left(\mathscr{R}-\frac{1}{2} g^{\mu\nu}\partial_\mu \phi \partial_\nu \phi - V \right) - 2 \int \dd^3 x \, \sqrt{h} \, \mathcal{K}\, ;\\
    \mathcal{S}_\text{b} & = - 2 \int \dd^3 x \, \sqrt{h} \, \mathcal{K}_0\, .
\end{split}
\end{equation}
The bulk action is supplemented by the usual boundary Gibbons-Hawking-York (GHY) term~\cite{York:1972sj,Gibbons:1976ue} depending on the extrinsic curvature $\mathcal{K}$, while $\mathcal{S}_\text{b}$ is the action of the background (with $\mathcal{K}_0$ the extrinsic curvature of its boundary) and it is introduced to make the Euclidean action well-defined and not divergent for spatially non-compact geometries. In the following, we will consider transitions between compact geometries only. The geometry of the Euclidean lump is given by
\begin{equation}
    \dd s^2 = U(r) \dd \tau^2 + U(r)^{-1} \dd r^2 + R(r)^2 \dd \Omega^2 \, ,
\end{equation}
where $\tau$ is the Euclidean time, $t = \text{i}\tau$, and $U(r)$ and $R(r)$ are given by \cref{Rlump,Usollump}, respectively. Given its dS asymptotics, its Euclidean manifold is compact. Just as in pure dS spacetime~\cite{Gibbons:1976ue}, thus, we do not need to support the bulk action with either the GHY term or the action of the background (thus, $\mathcal{S}_\text{b}= 0$). 

We now first evaluate the on-shell Euclidean action of the lump solution
\begin{equation}
    \mathcal{S}_\text{E} = -\int \dd^4 x \, \sqrt{g_\text{E}} \left(\mathscr{R}-\frac{1}{2} g^{\mu\nu}\partial_\mu \phi \partial_\nu \phi - V \right)\, .
\label{EuclideanActionLump}
\end{equation}
Exploiting \cref{EinsteinscalarEOM}, on shell, we have $\mathscr{R} = \partial_\alpha \phi \partial^\alpha \phi/2 + 2 V$. \Cref{EuclideanActionLump}, thus, yields
\begin{equation}
\begin{split}
    \mathcal{S}_\text{E} &= -\int \dd^4 x \, \sqrt{g_\text{E}} \, V = -4\pi \beta_\text{Lump} \int_0^{\rH} \dd r \, R^2(r) \, V(r)\, ,
\end{split}
\label{SEintegralpotential}
\end{equation}
where $\beta_\text{Lump}$ is the inverse of the temperature \eqref{temperaturelump}. Using \cref{Rlump,Vlump,rHell,temperaturelump}, we obtain 
\begin{equation}{\label{EuclideanActionLumpResults}}
    \mathcal{S}_\text{E}=-\frac{8 \sqrt{2} \left(\sqrt{5}+5\right)^{3/2} \pi^2 \ell^2}{5 \left(\sqrt{5}+1\right)} \simeq -134.329 \, \ell^2\, .
\end{equation}
We now compare it with the Euclidean action of the $\text{dS}_4$ vacua, which reads~\cite{Gibbons:1976ue}\footnote{In the following, we will not consider the $\text{dS}_2 \times \text{S}^2$ vacua. Their Euclidean action reads $\mathcal{S}_{\text{dS}_2 \times \text{S}^2}= 8\pi^2 L^2 > 0$, where $L$ is both the radius of the $2$-sphere and the dS length. Therefore, these vacua are thermodynamically less favored than both the lump and the pure dS spacetime (their probability amplitude is exponentially suppressed).}
\begin{equation}
    \mathcal{S}_\text{dS} = -16\pi^2 L^2 = -\frac{96\pi^2}{V(\phi_i)}\, ,
\end{equation}
where, in the second step, we have exploited the relation $V(\phi_i)/6 = L^{-2}$ (see below \cref{GRsolutions}). For the vacua \eqref{dSVacua}, one has
\begin{align}{\label{EuclideanActionExtremum}}
    \phi_0 &= 0 \, ,  &\mathcal{S}_\text{dS, 0} = -48\pi^2 \, \ell^2 \simeq -473.741 \, \ell^2\, ;\nonumber\\ 
    \phi_1 & \simeq 0.430664\, ,  &\mathcal{S}_\text{dS, 1} \simeq -491.596\, \ell^2 \, ; \nonumber \\
    \phi_2 & \simeq 2.16417\, , & \mathcal{S}_\text{dS, 2} \simeq -126.843 \, \ell^2\, .
\end{align}
From these results, it is evident that the probability amplitude to generate the lump is lower than that of the two $\text{dS}_4$ vacua $\phi_0$ and $\phi_1$, but higher than that of the absolute maximum $\phi_2$ (see \cref{Fig:VphiLump}). The latter is, thus, less preferred than the full lump solution, consistently with the tachyonic instability at the maximum. 
Through \cref{EuclideanActionPartitionFunction}, the hierarchy of free energies is as follows: $F_2 > F_\text{Lump} > F_0 > F_1$, where $F_{0, 1, 2}$ represent the free energies of the $\text{dS}_4$ vacua.  
The preferred and most stable configuration, that is, the one with the highest probability amplitude (lowest free energy), is the $\text{dS}_4$ vacuum corresponding to the minimum $\phi_1$. This mechanism provides the $\text{AdS}_4$ spacetime in the core with a channel to decay into a $\text{dS}_4$ solution. Thus, this process can be interpreted as the nucleation of $\text{dS}_4$ from $\text{AdS}_4$. These results complement those in Ref.~\cite{Kanno:2011vm}. While their work employs a mathematical approach closely aligned with ours, it focuses on transitions of the type $\text{dS} \to \text{AdS}$ or $\text{AdS} \to \text{AdS}$, without addressing the $\text{AdS} \to \text{dS}$ transitions that we explore.\\

We finally discuss the two limits $\ell \to 0$ and $\ell \to \infty$, already outlined at the end of \cref{sec:LumpSol}, in light of the Euclidean action results. 

For $\ell \to 0$, all Euclidean actions \eqref{EuclideanActionExtremum} evaluated at the stationary points of the potential go to zero, and hence the corresponding transition probability amplitudes become equivalent. This is due to the fact that, as detailed in \cref{sec:LumpSol}, $\ell \to 0$ corresponds to recovering a $\text{dS}_2 \times \text{S}^2$ from \cref{lumpfullsolution}. Therefore, there are no more extrema to transition to. The solution does not decay (there is no tachyonic instability) and stays in the $\text{dS}_2 \times \text{S}^2$ vacuum. 

Conversely, for $\ell \to \infty$, the geometry of the lump reduces to Minkowski spacetime, which is stable and does not decay. So the transition amplitudes loose their meaning. Considering directly the limit of \cref{EuclideanActionExtremum} would imply a divergent transition probability amplitude. However, we cannot perform smoothly the limit, since \cref{EuclideanActionExtremum} has been obtained considering a compact geometry, which Minkowski spacetime is not. In this case, we need to complement the bulk action with boundary terms. The total action is zero. Again, the probability amplitudes are all the same, implying that there is no actually another vacuum to transition to.

\subsection{Physical interpretation of the transition}

Let us now provide further physical insights into the transition process. Our analysis focuses on the behavior of the theory around the potential maximum $\phi_2$, where key information about the instanton mediating the transition can be extracted. Specifically, we aim to determine whether our instanton corresponds to a CDL or HM type.

The existence of a CDL bounce depends on the interplay between two critical factors: the absolute value of the scalar field mass at the potential barrier maximum, i.e., $|m^2_{\delta\phi}| \equiv |m^2_\text{dS}| \sim |\partial^2_\phi V(\phi_2)|$, and the curvature scale of the corresponding dS spacetime, $L_{\text{dS}}^{2} = 6/V(\phi_2)$~\cite{Jensen:1983ac}. When $|m^2_\text{dS}| \gg V(\phi_2)$, the thin-wall approximation applies: the CDL instanton exists and is degenerate with the HM one. Conversely, for $|m^2_\text{dS}| \sim V(\phi_2)$ or $|m^2_\text{dS}| \ll V(\phi_2)$, the interior of the bubble disappears and the Euclidean solution matches everywhere to the one at the potential maximum, resulting in an HM instanton. Geometrically, this means that the potential at the maximum flattens, invalidating the thin-wall approximation. 

To quantify these conditions, one usually considers the following ratios
\begin{equation}
    \alpha \equiv \frac{|\partial^2_\phi V(\phi_2)|}{H^2} \,, \qquad H^2 = \frac{V(\phi_2)}{6}\, .
    \label{betadef}
\end{equation}
$\alpha \gg 1$ ($\alpha >4$ for a wide class of potentials) corresponds to the existence of a CDL instanton, while $\alpha \ll 1$ ($\alpha < 4$) precludes it. We can also interpret these conditions in terms of the scale involved in the tachyonic instability at the maximum. As seen at the beginning of the previous section, the inverse of the effective mass of the scalar determines the time-scale of the instability $\tau_\text{inst}$. Therefore, a CDL instanton mediates the transition whenever $\tau_\text{inst}$ is shorter than the Hubble timescale $1/H$, or, equivalently, when the dS length becomes larger than the tachyon fluctuation scale. 

The interpretation of the transition is, however, quite intricate, as it crucially depends on whether the instanton scalar field oscillates between the true and false vacua or not. This can be studied by analyzing linear perturbations around the dS spacetime at the potential maximum, described by \cref{deltaphiequationofmotion}. Usually, global dS coordinates are employed to perform this study. It was shown in Ref.~\cite{Gregory:2020cvy}, however, that equivalent results can be obtained in the static patch by appropriately adjusting the field boundary conditions. It is found that the solution  $\delta \phi$ can feature a maximum number of nodes $N$, which is equal to the greatest integer satisfying the bound~\cite{Hackworth:2004xb,Lavrelashvili:2006cv,Battarra:2012vu}
\begin{equation}
    N(N+3) < \alpha\, .
    \label{boundbeta}
\end{equation}
In a dS background, the number of nodes corresponds to the number of negative eigenvalues~\cite{Lavrelashvili:2006cv}. When $N = 1$ (typical of a CDL bounce~\cite{Lavrelashvili:1999sr,Khvedelidze:2000cp}), we have a clear interpretation in terms of the instanton mediating the tunneling from the false to the true vacuum. When $N > 1$, the field oscillates back and forth between the true and false vacua. In this case, the interpretation is less straightforward. It is widely believed that such instantons do not represent transitions between vacua, but rather unstable and intermediate configurations~\cite{Lavrelashvili:2006cv,Brown:2007sd,Battarra:2012vu,Gregory:2020cvy}, whose role in the transition may be questioned. \\

We now apply this framework to our lump solution. Using \cref{potentialphi,dSVacua} into \cref{betadef} yields $\alpha \simeq 13 \gg 1$, indicating that the lump instanton at the potential maximum resembles a CDL instanton. However, we highlight some differences. Firstly, unlike standard CDL treatments, we consider a static configuration, neglecting the detailed time evolution during bubble nucleation. This approximation is justified by the lump's long instability timescale, comparable to the Hubble time, allowing us to focus on a static observer's perspective. Here, each constant-time slice represents a phase of the physical system. As a consequence, our global solution does not exhibit the conventional CDL instanton $O(4)$ symmetry, but rather this symmetry is broken to $O(3) \times U(1)$ by the presence of the lump. In our case, the $O(4)$ symmetry is preserved locally at the potential extrema, where the vacuum solution corresponds to dS spacetime. Fluctuations around the maximum $\delta \phi = \phi - \phi_2$, thus, describe deviations from this symmetry. Secondly, in our case, there is no exact vacuum to exact vacuum transition. Our system starts in a configuration corresponding to an approximate AdS spacetime, which then transitions to the true exact $\text{dS}_4$ vacuum. 

Another key ingredient in our analysis and in the comparison with the CDL instanton is assessing the number of negative modes of our lump solution. The latter usually depends on how many times the scalar-field solution crosses the $x$-axis during the transition between false and true vacuum. A key difference in our case is that the scalar field is a positive-definite function of $r$ (see \cref{scalarfieldlump}). A geometrically-equivalent way to assess the number of nodes in this case is counting the number of inflection points. Considering the whole range $r \in (-\infty, \infty)$, the scalar-field profile has two nodes, in agreements, therefore, with the general condition \eqref{boundbeta} with $\alpha \simeq 13$. However, in obtaining $V(\phi)$ in \cref{potentialphi}, we had to restrict to the interval $r \in [0,\infty)$, which reduces the number of nodes from $N = 2$ to $N = 1$ (in other words, in our framework, \cref{boundbeta} translates to $N(N+3) < \alpha/2$).  

Summarizing, our lump solution can be interpreted as a CDL instanton, mediating the transition between an approximately $\text{AdS}$ false vacuum to a true stable $\text{dS}_4$ one, through a tachyonic instability at the potential maximum. Here, the scalar field in the static patch of dS can be regarded as a thermal system at a temperature given by the corresponding dS spacetime at the maximum. Therefore, our bounce represents a \emph{thermal} CDL instanton mediating the transition, where the tachyonic instability is caused by thermal fluctuations at the maximum. This mechanism is akin to the one described in Ref.~\cite{Brown:2007sd}.

\section{Conclusions}
\label{sect:conclusions}

De Sitter spacetime is crucial for understanding the current accelerated expansion of the universe and the inflationary phase, both of which may require a QG framework. Motivated by the challenges of accommodating a dS vacuum in most approaches to QG and by the relevance of scalar fields in both black-hole physics and inflation, this paper explores the possibility of dS spacetime emerging from the AdS one in the framework of Einstein-scalar gravity. 

We specifically investigated a regular lump solution interpolating between an $\text{AdS}_4$ spacetime endowed with a linear scalar at $r = 0$ and a $\text{dS}_2 \times \text{S}^2$ spacetime at $r \to \infty$. Although we focused on a specific model, our results are expected to apply to a broader class of Einstein-scalar models exhibiting similar behavior.

Our analysis revealed that the $\text{AdS}_4$ in the core cannot be an exact vacuum of the theory, as it does not correspond to a quadratic stationary point of the potential. Instead, it serves as an approximate solution, sourced by a linear potential. We argue that the lack of an exact AdS vacuum in the core might be the origin of the model inherent instability. Remarkably, the scalar potential allowing for the lump solution includes several stationary points, with one being an absolute and stable minimum, corresponding to a $\text{dS}_4$ solution. Using the Euclidean action approach, we showed that this minimum is the preferred configuration. Due to its instability, the lump can be interpreted as a metastable state that eventually decays into the true, stable $\text{dS}_4$ vacuum. 

We suggest that our findings could potentially contribute to the still ongoing discussion regarding the compatibility of dS vacua within QG frameworks~\cite{Witten:2001kn,Dvali:2014gua,Dvali:2017eba,Dvali:2020etd}, such as supergravity and superstring theory. It is well known  that these theories naturally accommodate AdS vacua, while they  struggle to support stable (or even metastable) dS vacua without tachyonic excitations~\cite{Maldacena:2000mw,Maloney:2002rr,Townsend:2003qv,Kachru:2003aw,Burgess:2003ic,Balasubramanian:2005zx,Westphal:2006tn,Hertzberg:2007wc,Caviezel:2008tf,deCarlos:2009fq,Shiu:2011zt,Green:2011cn,Bena:2014jaa,Kutasov:2015eba,Junghans:2016uvg,Andriot:2016xvq,Moritz:2017xto,Andriot:2018ept,Cicoli:2018kdo,Obied:2018sgi,Danielsson:2018ztv,Andriot:2019wrs,Dine:2020vmr,Berglund:2022qsb,Cicoli:2024yqh,McAllister:2024lnt}. By demonstrating the possibility of metastable states transitioning from AdS to stable dS vacua, we have gained new insights into the landscape of possible solutions and their stability properties, while also presenting a promising way to address the aforementioned issues. Apart from fundamental QG considerations, our results may also be relevant for inflationary scenarios, where the dS phase in the cosmic evolution is commonly described by scalar-tensor gravity theories~\cite{Guth:1980zm,Linde:2007fr}. An intriguing possibility is that the inflationary phase could originate from an AdS $\to$ dS transition. Additionally, our results may provide a better understanding of hyper-compact spherically-symmetric objects, which act as black-hole mimickers. Solutions of this kind are known in Einstein-scalar gravity~\cite{Franzin:2018pwz,Cadoni:2023wxa}. They are regular (namely, singularity-free) solutions, interpolating between an AdS or dS core and an asymptotically-flat spacetime. Similar behavior has been observed in two-dimensional dilatonic theories, where dS emerges from AdS fluctuations~\cite{Biasi:2022ktq,Ecker:2022vkr}. Furthermore, the presence of an (A)dS core acts as a regulator of the geometry of a black-hole interior by removing the central singularity and playing a crucial role in the existence of regular black holes~\cite{Hayward:2005gi,Ansoldi:2008jw,Bonanno:2000ep,Cadoni:2022chn,Akil:2022coa,Franzin:2018pwz}.  

\bibliography{refs}
\end{document}